\documentclass[twocolumn,showpacs,preprintnumbers,amsmath,amssymb]{revtex4-1}
\usepackage{amsfonts, amsmath, bm, bbm, graphicx, epsfig, epstopdf}
\usepackage{dcolumn}
\usepackage{textcomp}
\begin{document}


\title{Photon-photon gate via the interaction between two collective Rydberg excitations}

\author{Mohammadsadegh Khazali, Khabat Heshami and Christoph Simon}
\affiliation{Institute for Quantum Science and Technology and Department of Physics
and Astronomy, University of Calgary, Calgary T2N 1N4, Alberta, Canada}

\date{\today}

\begin{abstract}
We propose a scheme for a deterministic controlled-phase gate between two photons based on the strong interaction between two stationary collective Rydberg excitations in an atomic ensemble. The distance-dependent character of the interaction causes both a momentum displacement of the collective excitations and unwanted entanglement between them. We show that these effects can be overcome by swapping the collective excitations in space and by optimizing the geometry, resulting in a photon-photon gate with high fidelity and efficiency.
 \end{abstract}

\maketitle

Deterministic quantum gates between individual photons are very desirable for photonic quantum information processing \cite{PhotonGate-1,PhotonGate-2,PhotonGate-5,PhotonGate-6,PhotonGate-4}. As photons interact only very weakly in free space, the implementation of such gates requires appropriate media. One attractive approach involves converting the photons into atomic excitations in highly excited Rydberg states, which exhibit strong interactions. Rydberg state based quantum gates between individual atoms and between atomic ensembles
have been proposed \cite{NutralAtomGateProposal-11,NutralAtomGateProposal-1,NutralAtomGateProposal-2,
NutralAtomGateProposal-3,NutralAtomGateProposal-4} and implemented \cite{NutralAtomGateImplement-1,NutralAtomGateImplement-2,NutralAtomGateImplement-3}. There are two categories of gates, those relying on the interaction between two excited atoms \cite{NutralAtomGateProposal-11,NutralAtomGateProposal-1}, and those based on Rydberg blockade \cite{NutralAtomGateProposal-1,NutralAtomGateProposal-2,
NutralAtomGateProposal-3,NutralAtomGateProposal-4,NutralAtomGateImplement-1,
NutralAtomGateImplement-2,NutralAtomGateImplement-3}, where only one atom is excited at any given time.
There is a significant body of work studying the effects of mapping photons onto collective atomic Rydberg excitations \cite{PhotAtomIntera-1,PhotAtomIntera-2,PhotAtomIntera-3,PhotAtomIntera-4,PhotAtomIntera-5,PhotAtomIntera-6,many-body}.
Most proposals for photon-photon gates involve propagating Rydberg excitations (polaritons), either using blockade \cite{PhotBlock-1,PhotBlock-2} or two excitations \cite{Phot-Int1,Phot-Int2,Phot-Int3,Phot-int4}.

Separating the interaction process and propagation makes it easier to achieve high fidelities for these photonic gates \cite{AdamStored}. Such separation can be achieved by photon storage, i.e. by converting the photons into stationary rather than moving atomic excitations. The only storage-based photonic gate that has been proposed so far is based on the blockade effect \cite{AdamStored}. Achieving blockade conditions can be challenging since both photons have to be localized within the blockade volume. Following \cite{NutralAtomGateProposal-11,NutralAtomGateProposal-1,Phot-Int1,Phot-Int2,Phot-Int3,Phot-int4}, we here propose a storage-based scheme that instead relies on the interaction between two stationary Rydberg excitations.

The main challenge for two-excitation based Rydberg gates in atomic ensembles arises from the fact that the interaction is strongly distance-dependent and thus not uniform over the profiles of the
two stored photons. We show that this a priori reduces the gate's fidelity by displacing the collective excitations in momentum space and by entangling their quantum states. However, we then show that it is possible to completely compensate the first effect by swapping the collective excitations in the middle of the interaction time, and to greatly alleviate the second effect by optimizing the shape and separation of the excitations, resulting in a photon-photon gate that achieves both high fidelity and high efficiency.

\begin{figure}
\scalebox{0.4}{\includegraphics*{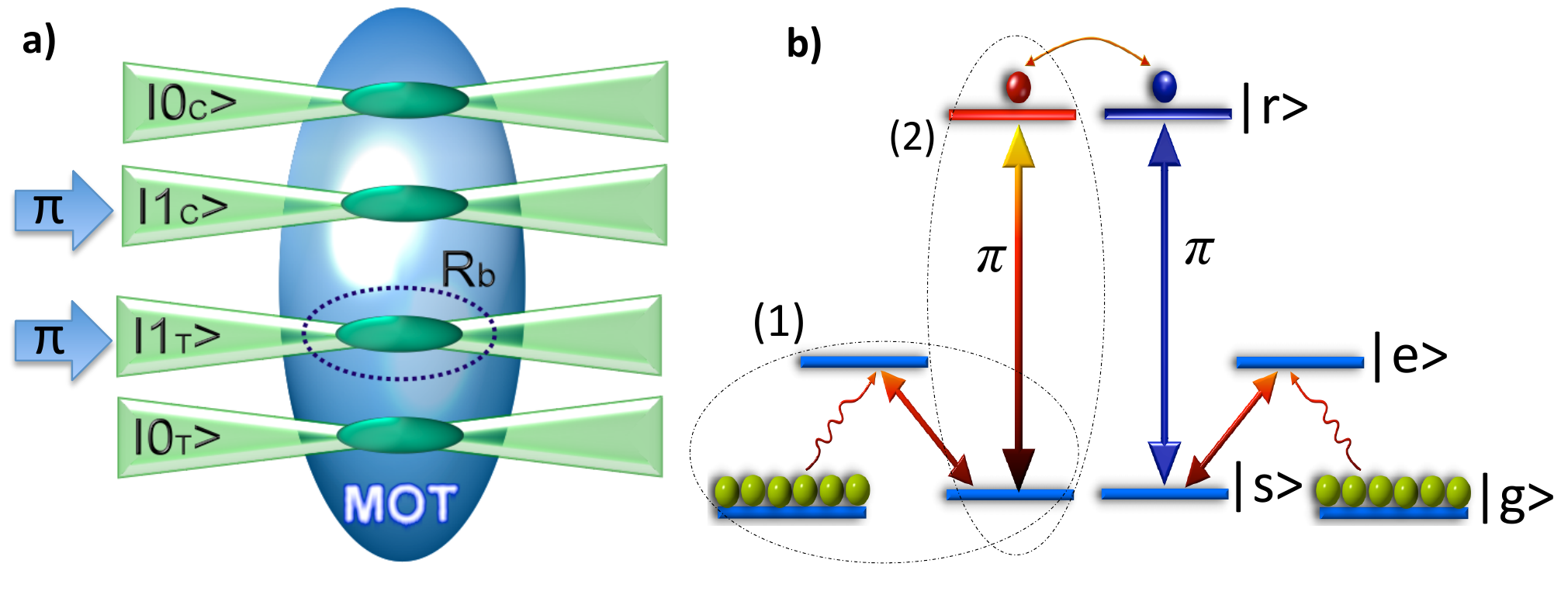}}
\caption{ (Color online) Proposed photon-photon gate scheme. (a) Setup. The scheme is based on dual
rail qubits \cite{ChuangYamamoto}. All four rails are stored as collective spin excitations in an atomic ensemble in a magneto-optical trap (MOT). Only the interacting
rails ($\left|1\right\rangle _{C}$ and $\left|1\right\rangle _{T}$)
are excited to Rydberg levels. The separation between the interacting rails is set to be larger than the blockade radius $R_b$ to ensure that both collective excitations can be promoted to the Rydberg level. (b) Level scheme. The photons are stored and retrieved through non-Rydberg EIT (dashed circle \#1), which completely separates the Rydberg interaction from the storage and retrieval process. Subsequently optical $\pi$ pulses promote the collective excitations in the interacting rails to Rydberg states (dashed circle \#2), where the van der Waals interaction creates a cumulative conditional phase. After the interaction time, the photons are retrieved by another pair of $\pi$ pulses followed by non-Rydberg EIT readout.} \label{Scheme}
\end{figure}

Now we describe our scheme in detail. As shown in Fig.~\ref{Scheme}a, information is encoded in dual-rail qubits \cite{ChuangYamamoto},
where the computational basis ($\left|0\right\rangle ,\left|1\right\rangle $)
is defined by two spatially separated paths. To implement a conditional
phase gate between control (C) and target (T) qubits, we store all
four rails in a cold alkaline atomic gas.  All four rails are stored and retrieved through non-Rydberg EIT in a lambda configuration (see Fig.~\ref{Scheme}b - Circle 1), which completely decouples the Rydberg interaction from the propagation. In comparison, the scheme of Ref. \cite{AdamStored} relies on Rydberg EIT (i.e. a ladder system involving a Rydberg state), such that the propagating polaritons are still interacting, albeit less strongly than the stored excitations.

The truth table for a controlled phase gate (with a controlled phase of $\pi$) is
$\left|a_{C}\right\rangle \left|a_{T}\right\rangle \rightarrow e^{i\pi a_{C}a_{T}}\left|a_{C}\right\rangle \left|a_{T}\right\rangle $, where $a\,\epsilon\,\{0,1\}$ and the phase is created under the condition that both photons are in the interacting rails $(\left|1\right\rangle _{C},\left|1\right\rangle _{T})$. Therefore, we only excite the interacting rails to Rydberg levels through optical $\pi$ pulses (see Fig.~\ref{Scheme}a, 1b-Circle 2), where the Rydberg interaction changes the energy of the interacting pair's state ($\left|1_{C}1_{T}\right\rangle$) and lets it accumulate a phase over time compared to the non-interacting pairs ($\left|1_{C}0_{T}\right\rangle
 ,\left|0_{C}1_{T}\right\rangle
 ,\left|0_{C}0_{T}\right\rangle$).
After the mentioned preparation steps, the wave function of the interacting
pair $(\left|1_{C}1_{T}\right\rangle) $ is given by
\begin{equation}\label{State}
\left|\Psi_{t_{0}}\right\rangle=\sum_{i1,i2}f_{i1}e^{-i\boldsymbol{k}_{10}.\boldsymbol{x}_{i1}}\hat{\sigma}_{r_{1}g}^{i1}f_{i2}e^{-i\boldsymbol{k}_{20}.\boldsymbol{x}_{i2}}\hat{\sigma}_{r_{2}g}^{i2}|G\rangle,
\end{equation}
where $\left|G\right\rangle=\otimes_{i=1}^{N}\left|g\right\rangle^{i}$ is the collective ground state and
$\boldsymbol{k_{10}}$ and $\boldsymbol{k_{20}}$ are the central wave vectors of the collective excitations. The summation in Eq. \ref{State} is over all atoms inside the medium. The raising operator $\hat{\sigma}_{r_{j}g}^{i}=|r_{j}\rangle^{i}\langle g|$ excites the $i$-th atom to the Rydberg state $|r_{j}\rangle$ ($j=1,2$). The spatial profile of the collective excitations is considered in $f_{i}$.
Their shape is determined through the storage process
and the shape of the input pulses. We assume a Gaussian profile for the rest of the paper.

The interaction energy between two Rydberg atoms has the form $\triangle(\boldsymbol{x})=\frac{c_{n}}{\left|\boldsymbol{x}\right|^{n}}$, where $\boldsymbol{x}$ is the separation between the atoms. It
changes from dipole-dipole $(n=3)$ in the short range to van~der~Waals $(n=6)$ in the long range \cite{NutralAtomGateImplement-1}. The spatial separation of the collective excitations in our protocol is in the range of the van~der~Waals interaction. The many-body interaction
Hamiltonian is $\hat{H}_{int}=\frac{1}{2}\sum_{_{l1\neq l2}}\hat{\sigma}_{r_{1}r_{1}}^{(l1)}\triangle(\boldsymbol{x}_{l1}-\boldsymbol{x}_{l2})\hat{\sigma}_{r_{2}r_{2}}^{(l2)}$,
where $\hat{\sigma}_{rr}$ is the projection operator.  Different combinations of excited atoms in Eq. (1) gain different phases, because their interaction strength is distance-dependent. This leads to a non-uniform distribution of conditional phase over each collective excitation, which affects the gate fidelity.

\begin{figure}[t]
\scalebox{0.48}{\includegraphics*{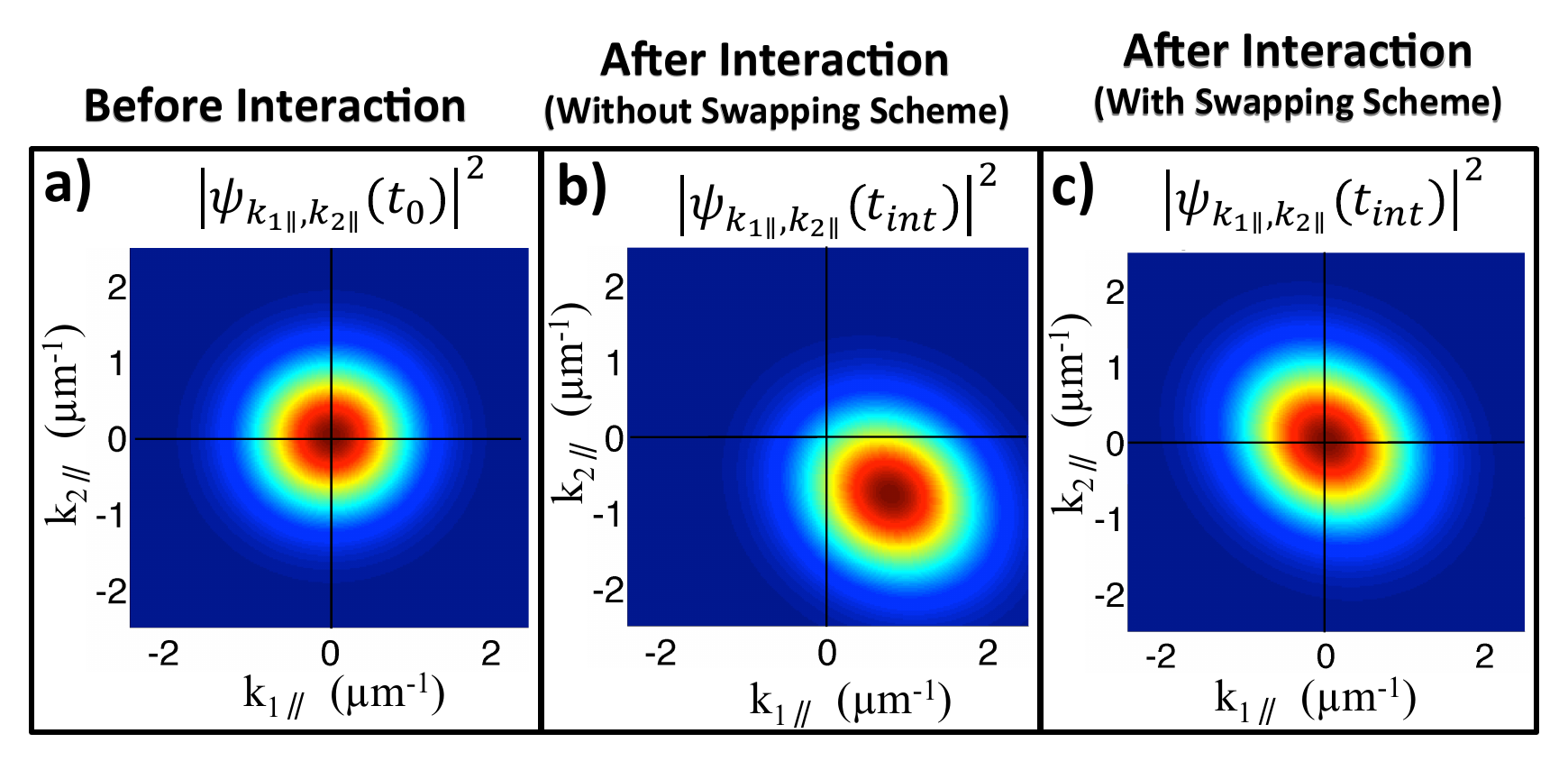}}
\caption{ (Color online) Two-excitation wave function in momentum space. Numerical evaluation of $\left|\psi_{k_{1\shortparallel},k_{2\shortparallel}}\right|^{2}$, for the momentum components $k_{1\shortparallel}$ and $k_{2\shortparallel}$ which are parallel to the separation between the collective excitations. (a) Before the interaction the two-excitation wave function is
a product of two individual Gaussian distributions. (b) After the interaction the center of the
distribution is displaced and its cross section becomes elliptic. The momentum displacement is created by the linear term of the interaction when expanded in terms of relative distance, see Eq. (3). The
elliptic shape is caused by the quadratic term in Eq. (3) and represents unwanted entanglement between the two excitations. (c) The displacement in momentum space can be compensated by a swapping protocol, see Fig. 3 and below. } \label{SiKy}
\end{figure}

In order to gain some insight into the dynamics of our system, we begin by deriving approximate analytic expressions for the effects of the non-uniform interaction. The modulus squared of the two-excitation wave function in momentum space, $\left|\Psi_{k_{1},k_{2}}(t)\right|^{2}$, is given by
\begin{equation} \label{Si2}
\left|\sum_{i1,i2}f_{i1}e^{-i\boldsymbol{K}_{1}.\boldsymbol{x}_{i1}}f_{i2}e^{-i\boldsymbol{K}_{2}.\boldsymbol{x}_{i2}}e^{\frac{-ic_{6}t}{\left|\boldsymbol{x}_{i1}-\boldsymbol{x}_{i2}\right|^{6}}}\right|^{2},
\end{equation}
where $\boldsymbol{K}_{j}=\boldsymbol{k}_{j}-\boldsymbol{k}_{j0}$ for $j=1$ ($j=2$) is the wave vector of the first (second) collective excitation relative to its central mode. When the collective excitations are far separated compared to their width, the interaction can be expanded to the second order in the relative distance,
\begin{eqnarray}\label{expansion}
&&\frac{1}{|\boldsymbol{x}_{i1}-\boldsymbol{x}_{i2}|^{6}}=\frac{1}{|\Delta \boldsymbol {x}_{0}|^{6}}-\frac{6(X_{i1\shortparallel}-X_{i2\shortparallel})}{|\Delta \boldsymbol{x}_{0}|^{7}}\\ \nonumber
&&-\frac{3(X_{i1\perp}-X_{i2\perp})^{2}}{|\Delta \boldsymbol{x}_{0}|^{8}}+\frac{21(X_{i1\shortparallel}-X_{i2\shortparallel})^{2}}{|\Delta \boldsymbol{x}_{0}|^{8}}+O(3),
\end{eqnarray}
where $\Delta\boldsymbol{x}_{0}=\boldsymbol{x}_{10}-\boldsymbol{x}_{20}$
is the distance between the center of the two Gaussian collective excitations and $\boldsymbol{X}_{i1}=\boldsymbol{x}_{i1}-\boldsymbol{x}_{10}$
indicates the relative position of an excited atom with respect to the center
of its distribution. The interaction can be separated into terms that are parallel and perpendicular with respect to the separation between the collective excitations $\Delta\boldsymbol{x}_{0}$, corresponding to the coordinates
$(\hat{x}_{\shortparallel}, \hat{x}_{\perp})$ etc.
One can correspondingly rewrite Eq.~(\ref{Si2}) in parallel and perpendicular dimensions, resulting in
\begin{align} \label{SiPP}
&\left|\psi_{k_{1\shortparallel},k_{2\shortparallel}}\right|^{2}\propto e^{\frac{w_{\shortparallel}^{2}}{2(1+4S_{\shortparallel}^{2})}\left[\left(K_{1\shortparallel}-k_{D}\right)^{2}+\left(K_{2\shortparallel}+k_{D}\right)^{2}+2S_{\shortparallel}^{2}\left(K_{1\shortparallel}+K_{2\shortparallel}\right)^{2}\right]} \nonumber \\
&\left|\psi_{k_{1\perp},k_{2\perp}}\right|^{2}\propto e^{\frac{-w_{\perp}^{2}}{2(1+4S_{\perp}^{2})}\left[K_{1\perp}^{2}+K_{2\perp}^{2}+2S_{\perp}^{2}\left(K_{1\perp}+K_{2\perp}\right)^{2}\right]},
\end{align}
where $2w_{\shortparallel}\,(2w_{\perp})$ is the spatial width of the collective excitation
in the parallel (perpendicular) dimension. The momentum displacement $k_{D}=\frac{6c_{6}t}{|\Delta \boldsymbol{x}_{0}|^{7}}$ is derived from the first order of the interaction expansion. The second order terms in  the parallel and perpendicular dimension give the coefficients $S_{\shortparallel}
=\frac{21w_{\shortparallel}^{2}c_{6}t}{|\Delta \boldsymbol{x}_{0}|^{8}}$ and $S_{\perp}=\frac{3w_{\perp}^{2}c_{6}t}{|\Delta \boldsymbol{x}_{0}|^{8}}$ respectively.

We numerically evaluate Eq.~(\ref{Si2}) and show the results  in Fig.~\ref{SiKy}(a,b) for the parallel dimension (see Fig.~5 in the supplemental materials \cite{supp} for the perpendicular dimension). These calculations are for the case where two co-propagating photons in the interacting rails are stored with a separation of $21\mu m$ in an ensemble of $^{87}Rb$ atoms in a MOT with a density of $\rho=4\times10^{12}$~cm$^{-3}$. Both collective excitations have the same spatial width $w_{\shortparallel}(w_{\perp})=3~\mu$m (8~$\mu$m), but they are excited to different
Rydberg levels $|103\, S_{1/2}\rangle$ and $|102\, S_{1/2}\rangle$ \cite{PhotAtomIntera-1}. Different principal numbers are considered for the two excitations in order to create a stronger interaction \cite{Rydberg-level-1,Rydberg-level-2}. The interaction time is 5 $\mu$s.

The numerical results correspond well to the expectations based on the approximate analytic treatment above. Fig. 2(b) clearly shows the expected displacement in momentum space, where the momentum shift $\boldsymbol{k}_{D}$ gained by the two collective excitations (in opposite direction and parallel to the separation) can be understood as being due to the action of the
Rydberg force $\boldsymbol{F}_{Ryd}=-\nabla U_{int}$ over the interaction time. In practice, this will result in retrieval of the photons in directions that deviate from the naively expected phase matching direction, see also the supplemental materials \cite{supp}. This ``frozen collision'' is a remarkable effect in the sense that the change of momentum due to the interaction only becomes apparent once the photons are read out. Based on the geometry of the valence orbital of excited atoms (which determines the sign of $c_6$), the collision can
be either attractive or repulsive \cite{Sign of interaction,Sign of interaction-2}.

Fig. 2(b) also shows the effect of the second-order term, which
creates unwanted entanglement between $\left|1\right\rangle _{C}$ and $\left|1\right\rangle _{T}$ (as well as spreading in momentum space). This term turns the circular cross section of the profile of the probability distribution in momentum space into a 45{\textdegree} rotated ellipse, see also Fig.~5(b) in the supplemental materials \cite{supp}. The relevant terms in the exponents in Eq. (4) are proportional to $e_{\shortparallel}^{2}=\frac{4S_{\shortparallel}^{2}}{1+4S_{\shortparallel}^{2}}$ and $e_{\perp}^{2}=\frac{4S_{\perp}^{2}}{1+4S_{\perp}^{2}}$, where $e_{\shortparallel}$ and $e_{\perp}$ are the eccentricities of the elliptic cross sections in the parallel and perpendicular dimension respectively.

\begin{figure}
\scalebox{0.4}{\includegraphics*{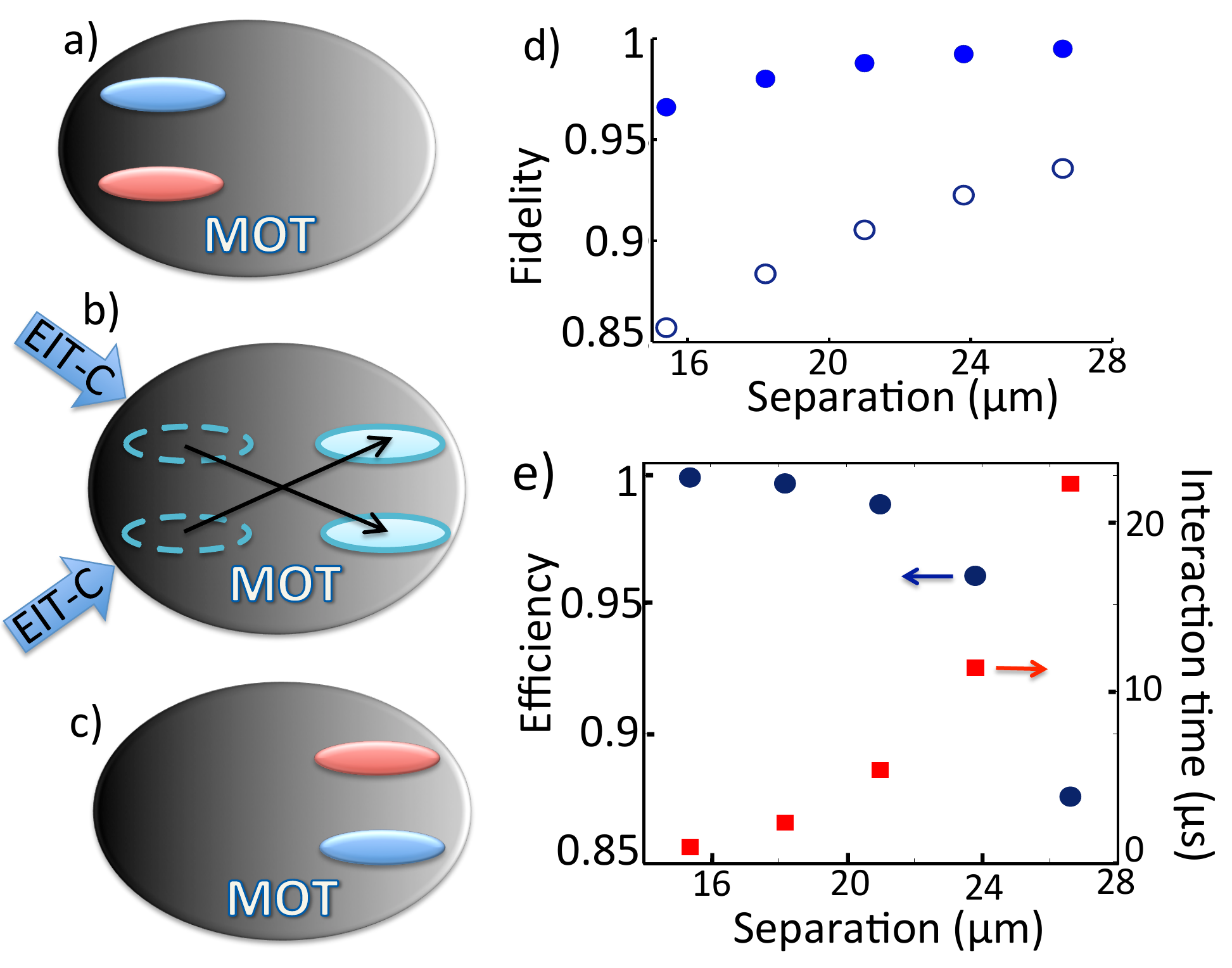}}
\caption{ (Color online) Swapping protocol to compensate the momentum displacement shown in Fig. 2(b) and resulting gate performance. (a) Photons in the interacting rails $\left(1_{C},\,1_{T}\right)$ are stored as collective excitations and excited to the Rydberg levels $|r\rangle$ as described in Fig.~\ref{Scheme}. They are brought back to the spin state $|s\rangle$ after half of the interaction time ($\frac{t}{2}$). (b) Tilted control
fields swap the relative positions of the two collective excitations  using non-Rydberg EIT. (c) The collective excitations are re-excited to the Rydberg levels, interact for $\frac{t}{2}$, and are de-excited again. The photons are retrieved using non Rydberg EIT.
(d) Gate fidelity as a function of the separation between the collective excitations. Solid and hollow circles are with and without the swapping protocol respectively. The spatial shape of the collective excitations is the same as in Fig. 2. (e) Gate efficiency (circles) and interaction time required for creating a $\pi$ phase shift (squares) as a function of the separation. The efficiency does not include photon storage and retrieval, see text. One sees that increasing the separation yields higher fidelity, but lower efficiency, because the weaker interaction for greater separations requires longer interaction times and hence more loss due to thermal motion and the finite lifetime of the Rydberg states. Using the swapping protocol, both high fidelity and high efficiency can be achieved.}\label{Swapping}
\end{figure}

We analyze the expected gate performance using the concepts of (conditional) fidelity and efficiency. Analogous concepts are commonly used in the context of quantum storage \cite{SimonEPJD}. The conditional fidelity quantifies the performance of the gate, conditioned on successful retrieval of both photons. The effects of photon loss are discussed in terms of efficiency below. Following the treatment in \cite{AdamStored}, the conditional fidelity of a gate operating on the initial state $\left|\phi\right\rangle =\frac{1}{2}\left(\left|0_{C}\right\rangle +\left|1_{C}\right\rangle \right)\left(\left|0_{T}\right\rangle +\left|1_{T}\right\rangle \right)$ can be quantified as $F=\sqrt{\left\langle \phi'\right|\rho\left|\phi'\right\rangle }$. This definition characterizes the gate's outcome $\rho$,
  relative to the ideal output  $\left|\phi'\right\rangle =(\left|00\right\rangle +\left|01\right\rangle +\left|10\right\rangle -\left|11\right\rangle )/2$. Since the many-body interaction only affects the interacting pair, the fidelity can be rewritten as
 $F=\sqrt{(9-3(\zeta+\zeta^{*})+\left|\zeta\right|^{2})/16}$ where $ \zeta=\left\langle \Psi_{t_{0}}\right|e^{-i\hat{H}_{int}t}\left|\Psi_{t_{0}}\right\rangle$ with $|\Psi_{t_0}\rangle$ as given in Eq. (1) and $\hat{H}_{int}$ as defined above. It is clear from Fig. 2(b) that the momentum displacement and the entanglement-related profile deformation will affect the value of $\zeta$ and hence of $F$. Controlling these effects is essential for achieving high gate fidelity.

\begin{figure}
\scalebox{0.33}{\includegraphics*{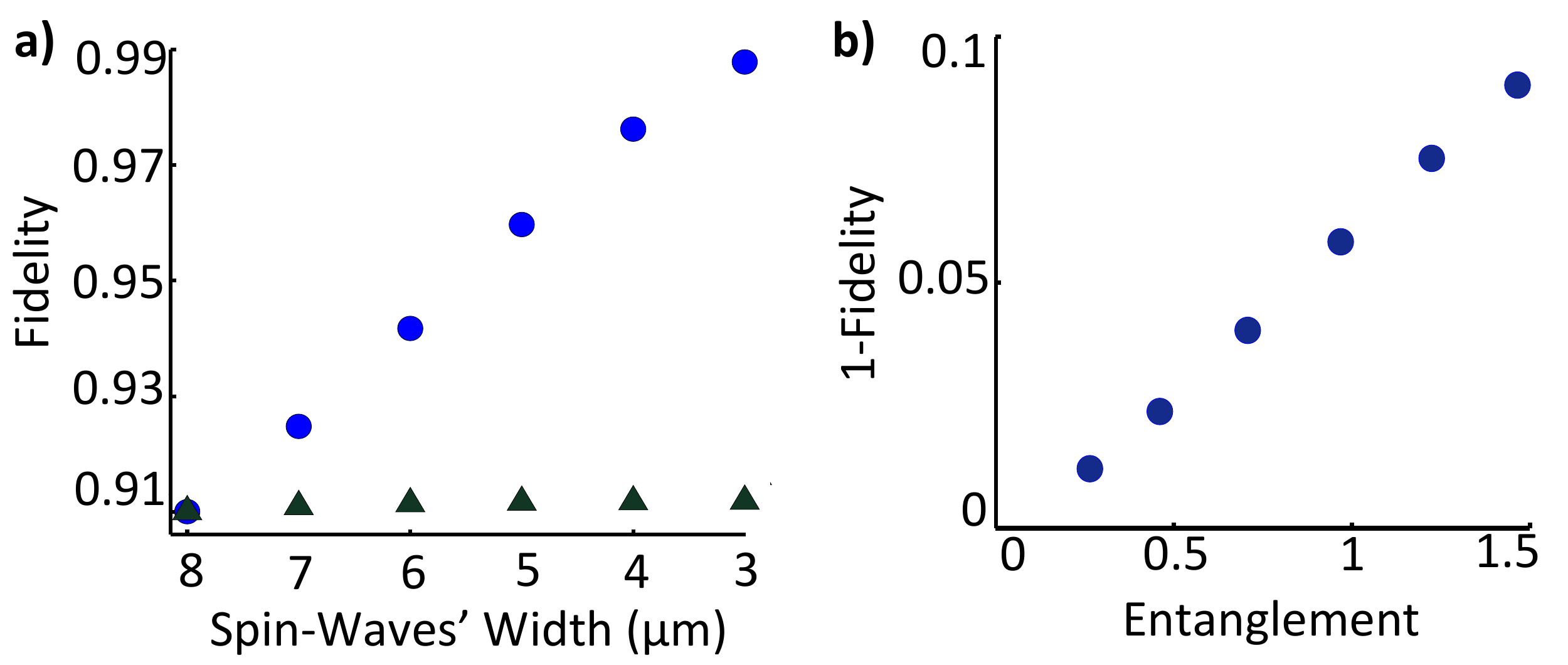}}
\caption{(Color online) Effects of unwanted entanglement on gate fidelity.  (a) The fidelity has a non-isotropic dependence on the width of the collective excitations. Here the collective excitations are separated by 21~$\mu$m and their spatial profile has the same initial width of 8~$\mu$m in all directions.
Compressing the width parallel to
the separation ($w_{\shortparallel}$) has a significant impact
on the fidelity (circles). In contrast, compressing the width perpendicular to the separation ($w_{\perp}$)
has a negligible effect (triangles).
(b) The fidelity reduction $1-F$ is proportional to the entanglement, quantified by the Von Neumann entropy. Here the momentum displacement is compensated by the swapping protocol of Fig. 3, leaving the unwanted entanglement as the main source of infidelity.}\label{Entanglement}
\end{figure}

We propose a swapping protocol to compensate the destructive effects of momentum displacement, see Fig. 3. Since the momentum displacement  $\boldsymbol{k}_{D}\propto\Delta\boldsymbol{x}_{0}t$ is proportional to the separation vector, swapping the relative position of the collective excitations $\left(\Delta\boldsymbol{x}_{0}\rightarrow-\Delta\boldsymbol{x}_{0}\right)$
in the middle of the interaction time reverses the rate of momentum displacement creation. The compensation of the momentum displacement after swapping can be seen in Fig. 2(c), and its beneficial effect on the gate fidelity in Fig. 3(d). The swapping protocol is relatively robust to positioning errors. In an example where the collective excitations are separated by 21~$\mu$m,  an averaged Gaussian error of 1~$\mu$m in the parallel dimension reduces the average fidelity by 1\%, see also the supplemental materials \cite{supp}. Errors in the perpendicular dimension are much less critical \cite{supp}.

It is important to also consider photon loss. Photon loss that is uniform over the four rails has no effect on the conditional fidelity as defined above. It can therefore be discussed independently in terms of the efficiency $\eta$, which is the probability of retrieving each photon after the gate operation. Non-uniform loss terms in our scheme can be made uniform by adding external sources of loss to certain rails, see supplementary information \cite{supp}. Two important sources of loss are atomic thermal motion \cite{supp,Thermal-motion} and the finite life time of the Rydberg levels (1.15 ms for $|102\, S_{1/2}\rangle$ and 1.18 ms for $|103\, S_{1/2}\rangle$) \cite{Life-time}. Their effects on the  efficiency are shown in Fig.~\ref{Swapping}(e) for different interaction times in an ensemble cooled to T=0.1~$\mu$K.
Considering the separation of interacting rails, there is a trade-off between fidelity and efficiency. A small separation improves the efficiency by reducing the interaction time (see Fig.~\ref{Swapping}(e)), but the resulting stronger interaction causes more entanglement and momentum displacement, which reduces the fidelity (see Fig.~\ref{Swapping}(d)). The swapping protocol makes it possible to achieve high fidelity and high efficiency simultaneously.

Another significant source of inefficiency comes from the process of storage and retrieval of single photons. A conservative estimate of the associated efficiency for the whole protocol (including the swapping) can be obtained by applying the photon's storage and retrieval efficiency twice \cite{Efficiency}. This corresponds to the use of two separate MOTs for storing photons before and after swapping. Based on this estimate the overall efficiency for a density of $\rho=4\times10^{12}$~cm$^{-3}$ (corresponding to an optical depth $d \approx 100$) is about 70\% . Increasing the density to $\rho=3.8\times 10^{13}$~cm$^{-3}$ ($d \approx 750$) improves the efficiency of repeated storage and retrieval to 95\%. In practice using a single MOT is likely to both be more practical and lead to higher efficiency than these estimates because in this case the stationary excitations only have to be converted into moving excitations (but not all the way into photons) at the intermediate stage.

We have shown how to compensate the effect of momentum displacement on the fidelity. The other destructive effect of the interaction that reduces the fidelity is the creation of unwanted entanglement between the collective excitations. Entanglement reduces the fidelity by deforming the two-excitation wave function in momentum space, see Fig.~\ref{SiKy}(b) (see also Fig.~5(b) in the supplemental materials \cite{supp}). Comparing the eccentricities of the ellipses in parallel and perpendicular direction obtained from Eq.~(\ref{SiPP}), $\frac{e_{\shortparallel}^{2}}{e_{\perp}^{2}}\backsim\frac{49w_{\shortparallel}^{4}}{w_{\perp}^{4}}$, one sees that the deformation is much stronger for the parallel dimension. Therefore, compression of the collective excitations parallel to their separation can reduce the unwanted entanglement while leaving room for extra atoms in the perpendicular dimensions in order to preserve the directionality of the collective emission
\cite{directionality1,directionality}.  The highly non-isotropic effects of profile compression on the fidelity are shown in Fig.~\ref{Entanglement}(a). The achievable width compression is mainly limited by diffraction. In order to show the relation between fidelity and entanglement even more clearly we calculate the Von Neumann entropy of the output state. Fig.~\ref{Entanglement}(b) shows that fidelity reduction and entanglement are indeed proportional.

In conclusion, we have proposed a photon-photon gate protocol that uses stationary collective Rydberg excitations, but does not rely on photon blockade. We have shown that unwanted effects due to the distance-dependence of the interaction are important but can be overcome, making it realistic to achieve a gate operation with high fidelity and efficiency.

{\it Acknowledgments.} We acknowledge financial support from AITF and NSERC. We thank B. He, A. Kuzmich, D. Paredes-Barato and B. Sanders for fruitful discussions.

\pagebreak

\widetext
\begin{center}
\textbf{\large Supplemental Materials for ``Photon-photon gate via the interaction between two collective Rydberg excitations''}
\end{center}

\setcounter{table}{0}
\setcounter{equation}{0}
\setcounter{page}{1}
\makeatletter
\renewcommand{\bibnumfmt}[1]{[SI#1]}
\renewcommand{\citenumfont}[1]{SI#1}

\title{Supplemental Materials for ``Photon-photon gate via the interaction between two collective Rydberg excitations''}

\author{Mohammadsadegh Khazali, Khabat Heshami and Christoph Simon}
\affiliation{Institute for Quantum Science and Technology
and Department of Physics and Astronomy, University of
Calgary, Calgary T2N 1N4, Alberta, Canada}

\date{\today}

\maketitle

\section{Effects of Interaction on the wave function in the perpendicular dimensions}

Fig. 5 shows the numerical evaluation of $\left|\psi_{k_{1\perp},k_{2\perp}}\right|^{2}$, the modulus squared of the two-excitation wave function in the dimensions perpendicular to the separation. While the second order of the interaction changes the Gaussian cross section from circular to elliptical by  entangling the two excitations, the first order does not have any effect in this dimension.

\begin{figure}[h]\label{SiPerpendicular}
\scalebox{0.5}{\includegraphics*{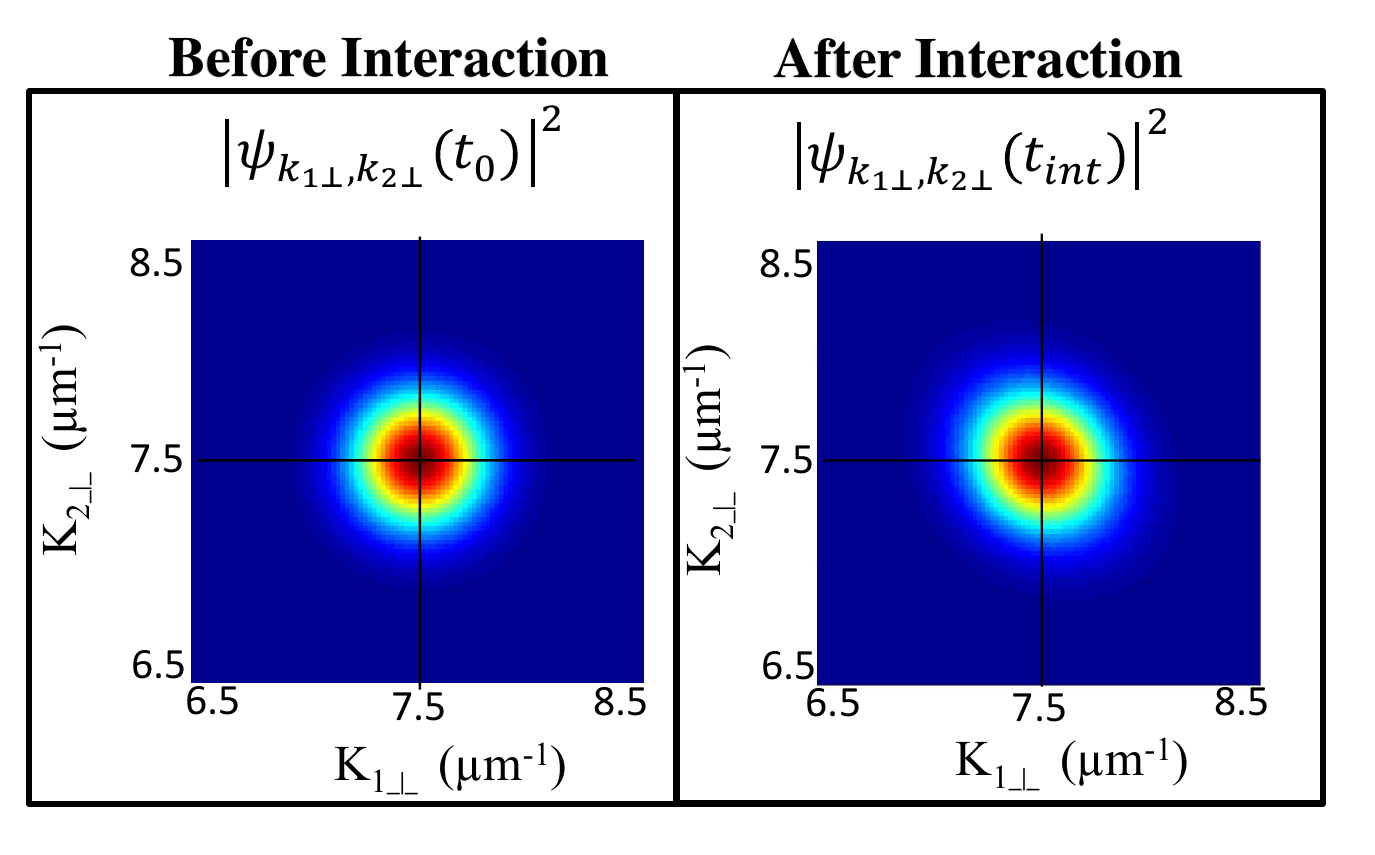}}
\caption{Numerical results for $\left|\psi_{k_{1\perp},k_{2\perp}}\right|^{2}$. There is no momentum displacement in this dimension. The parameters are the same as for Fig. 2.}
\end{figure}

\section{Sensitivity of Fidelity to positioning errors in Swapping}

Fig. 6 shows that the swapping protocol is more sensitive to positioning errors for the collective excitations in the parallel dimension than in the perpendicular dimension.

\begin{figure}\label{errorinswapping}
\scalebox{0.35}{\includegraphics*{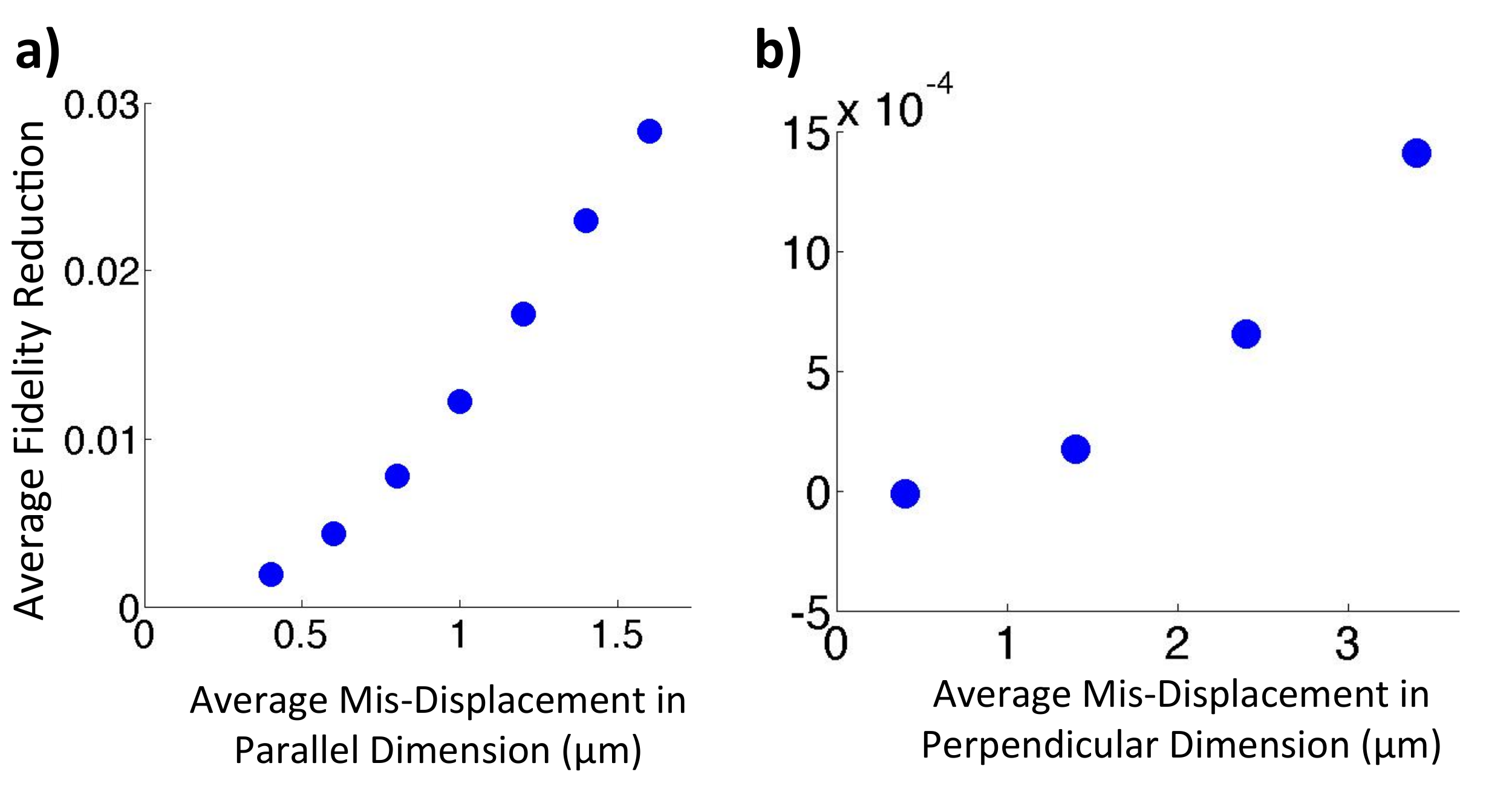}}
\caption{Average fidelity reduction as a function of positioning error for the swapping protocol in (a) parallel and (b) perpendicular dimension.
The parameters are the same as for Fig.~2.}
\end{figure}

\section{``Frozen Collision''}

Fig. 7 shows the redistribution of the momentum vectors of the collective excitations due to the interaction. One sees that collective excitations that are created by the storage of co-propagating photons will yield diverging photons upon retrieval.

\begin{figure}[h!]\label{AngularDestribution}
\scalebox{0.3}{\includegraphics{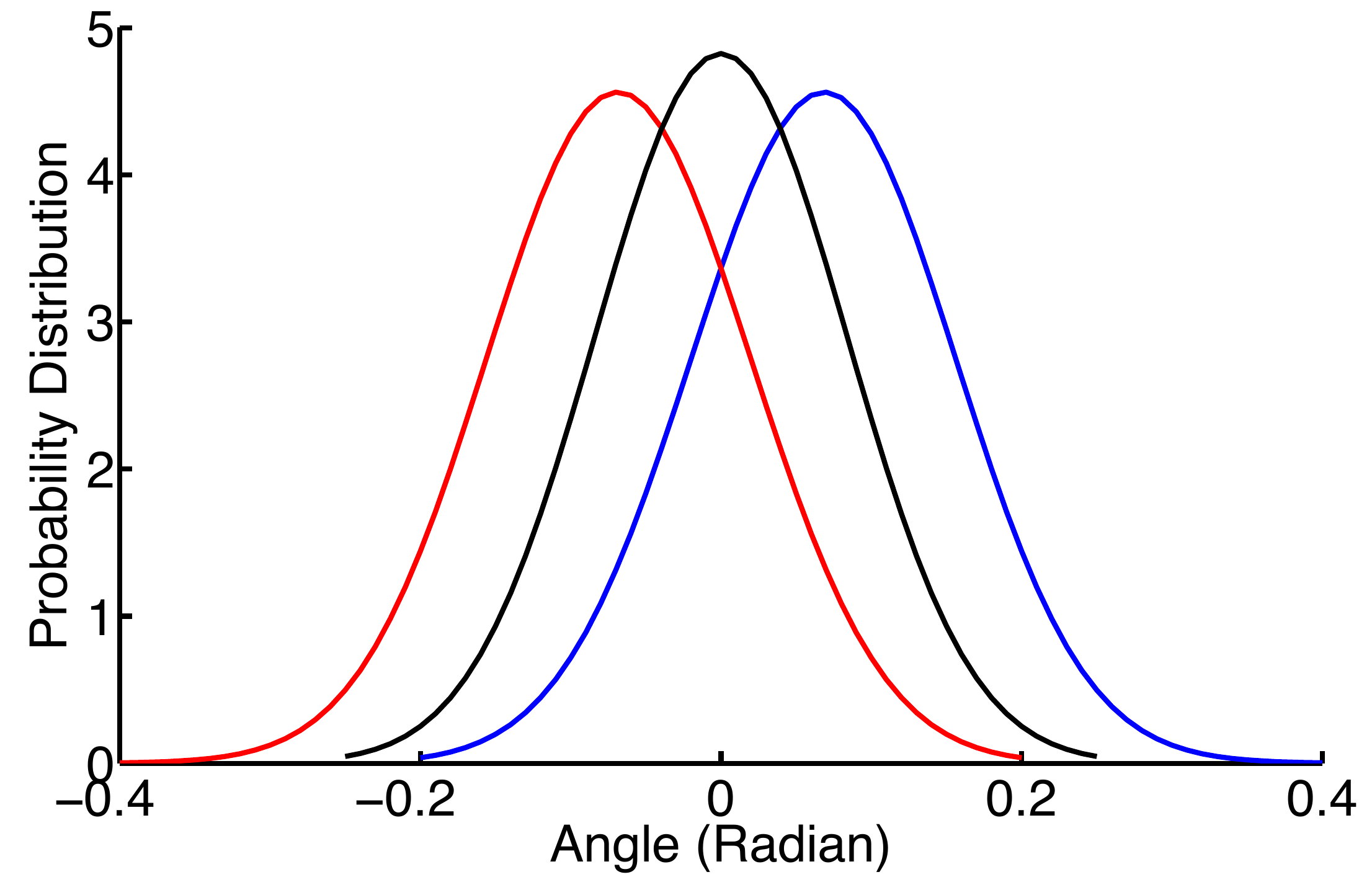}}
\caption{Angular distribution of the momenta of the collective excitations before (black for both) and after (red and blue) interaction. The parameters are the same as for Fig. 2.}
\end{figure}

\section{Gate Efficiency: thermal motion and uniform loss}

\subsection{Thermal motion}

The efficiency factor due to the thermal motion of the atoms is given by $\eta_{_{th}}=\frac{1}{(1+(\frac{t}{\xi})^{2})^{2}}exp[\frac{-t^{2}/\tau^{2}}{(1+(t/\xi)^{2})}]$
\cite{key-1}, where $\tau=\frac{\Lambda}{2\pi v}$ is the dephasing
time scale, which is determined by the wave length $\Lambda$ of the collective excitations and the thermal speed $v$, and $\xi=\frac{w}{v}$ is the time scale on which an atom traverses the width $w$ of a collective excitation.

\subsection{Uniform loss}

Photon loss that is uniform over the four rails has no effect on the conditional fidelity and can be quantified independently in terms of the efficiency. Since only the interacting rails ($\left|1_{C}\right\rangle ,\left|1_{T}\right\rangle $) are excited
to Rydberg levels, they experience an extra loss due to the life time of the Rydberg level. Furthermore, the shorter wavelength of the Rydberg excitations in the interacting rails creates a stronger loss due to the atomic thermal motion compared to the non-interacting rails. Finally, the extra process of storage and retrieval during the swapping of the interacting rails causes an extra loss of efficiency in these rails. The loss can be made uniform by adding a controlled external source of loss
on the non-interacting rails ($\left|0_{C}\right\rangle ,\left|0_{T}\right\rangle $).

\end{document}